\documentclass[twocolumn,showpacs]{revtex4}
\usepackage{graphicx}
\usepackage{dcolumn}
\usepackage{bm}
\usepackage{longtable}

\newcommand{\ima}{{\mbox{Im}\,}}

\begin{document}

\title{Quark mass dependence of $\rho$ and $\sigma$ Mesons
from Dispersion Relations 
\\and Chiral Perturbation Theory}

\author{ C. Hanhart$^a$,  J. R. Pel\'aez$^b$ and G.R\'ios$^b$ }

\address{$^a$ Institut f\"ur Kernphysik (Theorie), Forschungzentrum J\"ulich, D-52425 J\"ulich, Germany}
\address{$^b$ Dept. F\'{\i}sica Te\'orica II. Universidad Complutense, 28040, Madrid. Spain.}

\begin{abstract} 
  We use the one--loop Chiral Perturbation
  Theory $\pi\pi$--scattering amplitude and dispersion theory in the
  form of the inverse amplitude method to study the quark mass
  dependence of the two lightest resonances of the strong
  interactions, the $f_0(600)$ ($\sigma$) and the
  $\rho$ meson. As main results we find that the $\rho\pi\pi$
coupling constant is almost quark mass independent and that
the $\rho$ mass shows a smooth quark-mass dependence while
that of the $\sigma$ shows a strong nonanalyticity.
These findings are  important for studies of the meson
  spectrum on the lattice.
\end{abstract}
 
\maketitle

\vspace{0.8cm}
\newcommand{\boldpi}{\mbox{\boldmath $\pi$}}
\newcommand{\boldtau}{\mbox{\boldmath $\tau$}}
\newcommand{\boldT}{\mbox{\boldmath $T$}}
\newcommand{\gaprox}{$ {\raisebox{-.6ex}{{$\stackrel{\textstyle >}{\sim}$}}} $}
\newcommand{\saprox}{$ {\raisebox{-.6ex}{{$\stackrel{\textstyle <}{\sim}$}}} $}

Although studied theoretically as well as experimentally for
many years, the spectrum of the lightest resonances in QCD
is still not understood from first principles. The only known
way to extract nonperturbative quantities from QCD is the
use of lattice QCD. However, current calculations are typically still
done for relatively high quark masses (see, e.g., Refs.~\cite{Aoki:1999ff,scalars}). 
Thus, in order to make 
contact with experiment, appropriate extrapolation formulas need to be
derived. This is typically done by using chiral perturbation theory (ChPT),
the low energy effective theory of QCD~\cite{chpt1,chpt2}.
ChPT predictions are model independent and, in particular, provide, as
an expansion, the dependence of observables on the quark masses
(or equivalently the pion mass).  The aim of this Letter is to predict
the quark-mass dependence of the
$\sigma$ and the $\rho$ mesons from basic principles, namely, using
ChPT to next--to--leading order (NLO), unitarity, and 
analyticity in the form of dispersion theory using
the inverse amplitude method (IAM)~\cite{modIAM}. It is
obtained from a subtracted dispersion relation of the inverse amplitude, whose
imaginary part in the elastic region is known exactly from unitarity.
All dependences on QCD
parameters appear through the ChPT expansion, which is used to
calculate the low energy subtraction points and the left cut. Hence,
up to a given order in ChPT,
the approach has no model dependences. For this work, we will use the NLO $SU(2)$
elastic IAM whose parameters are fitted to $\pi\pi$ scattering data,
and therefore, our results are  {\it model independent only up to NLO ChPT.}

The use of dispersive methods also allows for a straightforward
extension to the second Riemann sheet of the complex plane where poles
associated to resonances occur. In this way both $\sigma$ and
$\rho$ poles appear naturally without any further assumptions and have
the correct dependence on QCD parameters up to the order of the ChPT
expansion used in the IAM.  For instance, from a study of the leading
$1/N_c$ behavior of the amplitudes it was possible to conclude that
the $\rho$ is mostly of $\bar qq$ nature whereas the $\sigma$ is
predominantly non--$\bar qq$~\cite{Pelaez:2006nj}.

In this Letter, we study how this different structure encoded in the ChPT
parameters gets reflected in the quark mass dependence.  As we will
see, also with respect to the quark-mass dependence, the $\rho$ behaves
differently than the $\sigma$. We will show that for sufficiently large
quark (pion) masses, both states become stable poles on the physical
sheet; however, this limit is approached very differently, 
only in part due to their different quantum numbers.

Another motivation for this study is the Anthropic Principle
\cite{AP}, i.e., the need for a subtle fine tuning of various
parameters of the Standard Model.  In order to allow for an efficient
triple $\alpha$ process, necessary for the production of carbon, a
$NN$ interaction within 2~\% of the known strength is necessary. Since
the $\sigma$ meson plays a central role in the $NN$ interaction, this
leads to bounds for the sigma mass and, correspondingly, for the
quark masses~\cite{Jeltema:1999na}.  This issue is
particularly exciting, 
if the fundamental constants were time dependent,
as claimed in Ref.~\cite{claim}.  In this context
the quark mass dependence of the $\sigma$ was studied in
Ref.~\cite{flammbaumshuryak}.
  
We are interested in $\pi\pi$ elastic amplitudes projected on
partial waves $t_{IJ}$ of definite isospin $I$ and total angular
momentum $J$.  For simplicity, we will drop the $IJ$ labels. Note
that the $\sigma$ ($\rho$) resonance appears as a pole in the second
Riemann sheet of the partial wave $(I,J)=(0,0)$ ($(1,1)$).
Elastic unitarity implies:
\begin{equation}
	\ima t(s)=\sigma(s)\vert t(s)\vert^2, \quad\Rightarrow\quad	\ima
 [{t(s)}^{-1}]=-\sigma(s),
	\label{unit}
\end{equation}
where $s$ is the usual Mandelstamm variable for the total energy,
$\sigma(s)=p/(2\sqrt{s})$ and $p$ is the center of mass momentum.
Consequently, the imaginary part of the elastic inverse amplitude is known exactly.

In this Letter, we focus on the two lightest resonances
of QCD, the $\rho (770)$ and the $f_0(600)$.
It is therefore enough to work with the two lightest quark
flavors $u,d$ in the isospin limit of an equal mass 
$\hat m=(m_u+m_d)/2$. The pion mass is given by an
expansion $m_\pi^2\sim \hat m+...$~\cite{chpt2}.
Therefore, studying the quark mass dependence is equivalent to studying
the $m_\pi$ dependence.

ChPT amplitudes are obtained as a series expansion $t=t_2+t_4+...$
with $t_k=O(p^k)$, where $p$ stands generically either for pion
momenta or masses.  The leading order, $t_2$, which is obtained at
tree level from the $O(p^2)$ Lagrangian, is just a polynomial fixed by
chiral symmetry in terms of the pion mass and its decay constant
$f_\pi$.  The NLO term $t_4$ has both one-loop contributions from the
$O(p^2)$ Lagrangian and tree level contributions from the $O(p^4)$
Lagrangian. The latter depend on a set of low energy constants (LEC),
denoted by $l_i$, that absorb the one-loop divergences through
renormalization.   Since these constants are the
coefficients of the energy and mass expansion {\it the $l_i$ have
no quark mass dependence}.  In contrast, $f_\pi$
gets renormalized at NLO and thus depends explicitly on the pion mass.
  
The ChPT series, being an expansion, satisfies
unitarity, Eq.(\ref{unit}), just perturbatively:
\begin{equation}
	\ima t_4(s)=\sigma(s)\vert t_2(s)\vert^2, \quad\Rightarrow\quad	\ima \frac{t_4(s)}{t_2(s)^2}=\sigma(s),
	\label{pertunit}
\end{equation}
and cannot generate poles. Therefore the resonance region lies beyond
the reach of standard ChPT.  However, it can be reached by combining
ChPT with dispersion theory either for the amplitude~\cite{gilberto}
or the inverse amplitude through the
IAM~\cite{Truong:1988zp,Dobado:1996ps,Guerrero:1998ei}.

{\bf 3.} 
The IAM
uses the ChPT
series to generate resonances in meson-meson scattering. In the
elastic case, the IAM follows from dispersion theory
\cite{Truong:1988zp,Dobado:1996ps} due to the fact that $t$ and $1/t$
have an almost identical analytic structure: both have a ``physical
cut'' from threshold to $\infty$ and a ``left cut'' from $-\infty$ to
$s=0$.  In addition, the inverse amplitude may have a pole whenever
the amplitude vanishes.  For the $\rho$--channel, this happens at
threshold to all orders, but for scalar waves it occurs at the so-called
Adler zero $s_A$, that lies on the real axis below threshold, thus
within the ChPT region of applicability.  Its position can be
obtained from the ChPT series, i.e., $s_A=s_2+s_4+...$, where $t_2$ vanishes at
$s_2$, $t_2+t_4$ at $s_2+s_4$, etc...
Thus we can write a dispersion
relation for the inverse amplitude,
\begin{equation}
  \label{1/tdispA}
  \frac1{t(s)}=\frac{s-s_A}{\pi}\int_{RC}dz\,
  \frac{\ima 1/t(z)}{(z-s_A)(z-s)}+
  LC_{\frac{1}{t}}+PC_{\frac{1}{t}},
\end{equation}
where ``LC'' stands for a similar integral over the left cut.
To ensure convergence we have made a subtraction precisely at $s_A$. 
 Since $t_2$ is real on the real axis, we can similarly write
\begin{equation}
  \label{t4/t2dispA}
  \frac{t_4(s)}{t_2(s)^2}=\frac{s-s_2}{\pi}\int_{RC}dz\,
  \frac{\ima t_4(z)/t_2(z)^2}{(z-s_2)(z-s)}+
  LC_{\frac{t_4}{t_2^2}}+PC_{\frac{t_4}{t_2^2}}.
\end{equation}
We can now use unitarity, Eqs.(\ref{unit}) and (\ref{pertunit}), to find that the 
integrand numerators 
are {\it exactly opposite}. 
Since the $LC$ integral is weighted at low energies, using NLO ChPT, we also find
 $LC(1/t)\simeq -LC(t_4/t_2^2)$. 
Consequently, crossing symmetry is satisfied just up to NLO.
In the above relations, $PC_{{1}/{t}}$ and $PC_{{t_4}/{t_2^2}}$ 
stand for the contributions of a double
and triple pole respectively, which can be easily calculated 
within ChPT. Their contribution is only relevant around the Adler zero, 
where they diverge. Finally, we can use ChPT to 
approximate $(s-s_A)/(z-s_A)\simeq (s-s_2)/(z-s_2)$.
Altogether, we find 
\begin{eqnarray}
t^{mIAM}(s)&=& \frac{t^2_2(s)}{ t_2(s)-t_4 (s)+A^{mIAM}(s)},\label{mIAM}\\
  \label{Aangelpipi}
A^{mIAM}(s)&=&t_4(s_2){-}\frac{(s_2{-}s_A)(s{-}s_2)\left[t'_2(s_2){-}t'_4(s_2)\right]}{s{-}s_A}. \nonumber
\end{eqnarray}
The usual IAM is recovered for $A^{mIAM}\equiv0$, which holds exactly
for all partial waves but the scalar ones.  In the original IAM
dispersive derivation \cite{Truong:1988zp,Dobado:1996ps} $A^{mIAM}$
was neglected, since it formally yields a NNLO contribution.  However,
due to neglecting $A^{mIAM}$, the IAM has a spurious pole close to its
Adler zero, which is only correct to LO ChPT.  As we will see, for
large $m_\pi$ the $\sigma$ pole splits into two virtual poles below
threshold, one of them moving towards zero and eventually approaching
the spurious pole of the IAM.  Thus, although the Adler zero ($\simeq
m_\pi/\sqrt{2}$) is very deep in the subthreshold region, we 
used the modified IAM (mIAM)
\cite{FernandezFraile:2007fv,modIAM}, which has no spurious pole, and
reproduces the Adler zero up to NLO.  Switching from the IAM
to the mIAM influences only the mentioned second $\sigma$ pole, and
only when it is very close to the spurious pole (this occurs when
$M_\sigma\leq 1.5 m_\pi$). Besides this, the IAM and mIAM results are
essentially the same. For
subtractions made at different low energy points, $A^{mIAM}$ acquires
additional NNLO terms, but their effect is negligible~\cite{modIAM}.

Therefore, {\it up to a given order in ChPT}, the elastic (m)IAM
is built in a model independent way from the first principles of
unitarity and analyticity in the form of a dispersion relation.
The ChPT series is used only on the Adler zeros and the 
left hand cut which is heavily weighted at low energies, 
thus well within its region of applicability. 
It is a dispersion integral for the inverse amplitude
that allows us to study the resonance region.

Although remarkably simple, the (m)IAM amplitudes satisfy elastic
unitarity, Eq.(\ref{unit}), exactly, and provide a very good
description of meson-meson scattering data simultaneously in the
resonance and low energy
regions~\cite{Truong:1988zp,Dobado:1996ps,Guerrero:1998ei}.
Furthermore, the IAM generates the poles in the second Riemann sheet
associated to the resonances, namely $\sigma$ and $\rho$.
This description is obtained with values of the LEC compatible with
those of standard ChPT \cite{Dobado:1996ps,Guerrero:1998ei}. Actually,
the ChPT series up to NLO is recovered when Eq. ({\ref{mIAM}}) is
reexpanded at low energies. Thus,
after a fit to data we can modify $m_\pi$ and follow the
poles associated to $\rho$ and $\sigma$ on the second sheet.

As long as they fall within their uncertainties, the precise
values of the LECs $l_3^r$ and $l_4^r$ are not very relevant for this
study.  We take from \cite{chpt2} $10^{3}l_3^r=0.8\pm 3.8,\,10^{3}l_4^r=6.2\pm
5.7$. Then we fit the mIAM to data up to the resonance region and find
$10^{3}l_1^r=-3.7\pm 0.2,\,10^{3}l_2^r=5.0\pm 0.4$. All these LEC are evaluated
at $\mu=0.77\,{\rm GeV}$. 

The values of $m_\pi$ considered should fall within the ChPT range of
applicability and allow for some elastic $\pi\pi$ regime to exist
below $K \bar K$ threshold.  Both criteria are satisfied, if $m_\pi
\leq 0.5\,{\rm GeV}$, since we know SU(3) ChPT still works fairly well
with such a kaon mass, and because for $m_\pi\simeq 0.5$ GeV, the kaon
mass becomes $\simeq$ 0.6 GeV, leaving a 0.2 GeV gap to the
two-kaon threshold. For larger values of $m_\pi$ a coupled-channel IAM is
needed, which is feasible, but lies beyond our present scope, and
lacks a dispersive derivation.

\begin{figure}
\begin{center}
\includegraphics[scale=0.25,angle=0]{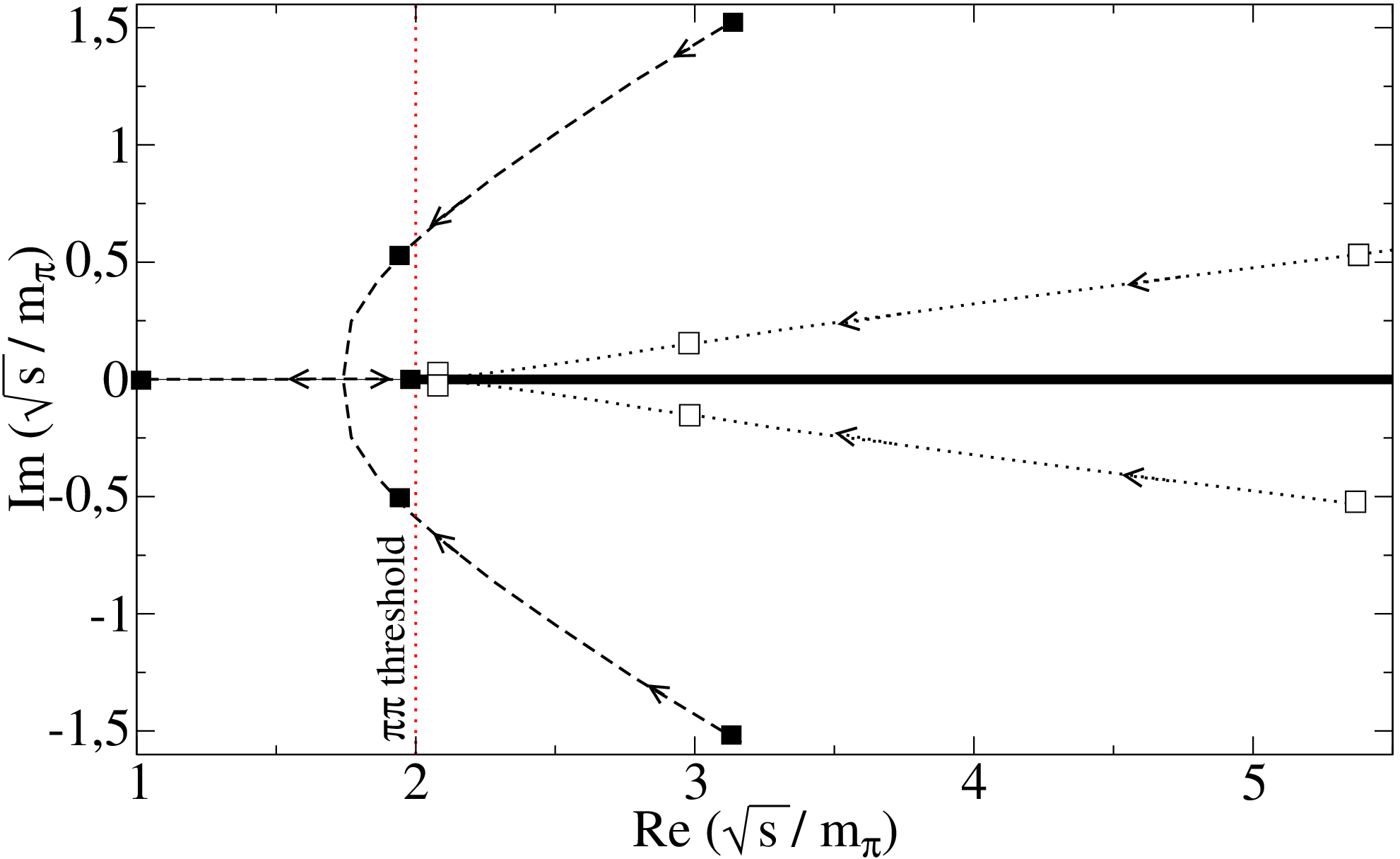}
\caption{ Movement of the $\sigma$ (dashed lines) and $\rho$ (dotted
  lines) poles for increasing pion masses (direction indicated by the
  arrows) on the second sheet.  The filled (open) boxes denote the
  pole positions for the $\sigma$ ($\rho$) at pion masses $m_\pi=1,\
  2,$ and $3 \times m_\pi^{\rm phys}$, respectively. Note, for
  $m_\pi=3m_\pi^{\rm phys}$ three poles accumulate in the plot very near the
  $\pi\pi$ threshold.}
\label{polosNormUpDown}
\end{center}
\end{figure}

Fig.~\ref{polosNormUpDown} shows, in the second Riemman
sheet, the $\rho $ and $\sigma$ poles for the physical $m_\pi$, and
how they move as $m_\pi$ increases. Note that, associated to each
resonance, there are two conjugate poles that move symmetrically on
each side of the real axis.  In order to see more clearly that all
poles move closer to the two-pion threshold, which is also increasing,
all quantities are given in units of $m_\pi$ so that the two-pion
threshold is fixed at $\sqrt{s}=2$. Let us recall that, for narrow
resonances, their mass $M$ and width $\Gamma$ are related to the pole
position in the lower half plane as $\sqrt{s_{pole}}\simeq M
-i\Gamma/2$ and customarily this notation is also kept for broader
resonances.  Hence, both $\Gamma_\sigma$ and $\Gamma_\rho$
decrease for increasing $m_\pi$.  In particular, $\Gamma_\rho$
vanishes exactly at threshold where one pole jumps into the first
sheet, thus becoming a traditional stable state, while its 
partner remains on the second sheet practically at the very same
position as the one in the first.  In contrast, when $M_\sigma$
reaches the two-pion threshold, its poles remain on the second sheet
with a non-zero imaginary part before they meet on the real axis and
become virtual states.  As $m_\pi$ increases further, one of those
virtual states moves towards threshold and jumps onto the first sheet,
whereas the other one remains in the second sheet.  Such an
analytic structure, with two very asymmetric poles in different sheets
of an angular momentum zero partial wave, is a strong indication for a
prominent molecular component \cite{Weinberg,baru}.  Differences between
P-wave and S-wave pole movements were also found within quark models
\cite{vanBeveren:2002gy}, the latter also showing two second sheet
poles on the real axis below threshold.

\begin{figure}
\includegraphics[scale=0.21,angle=0]{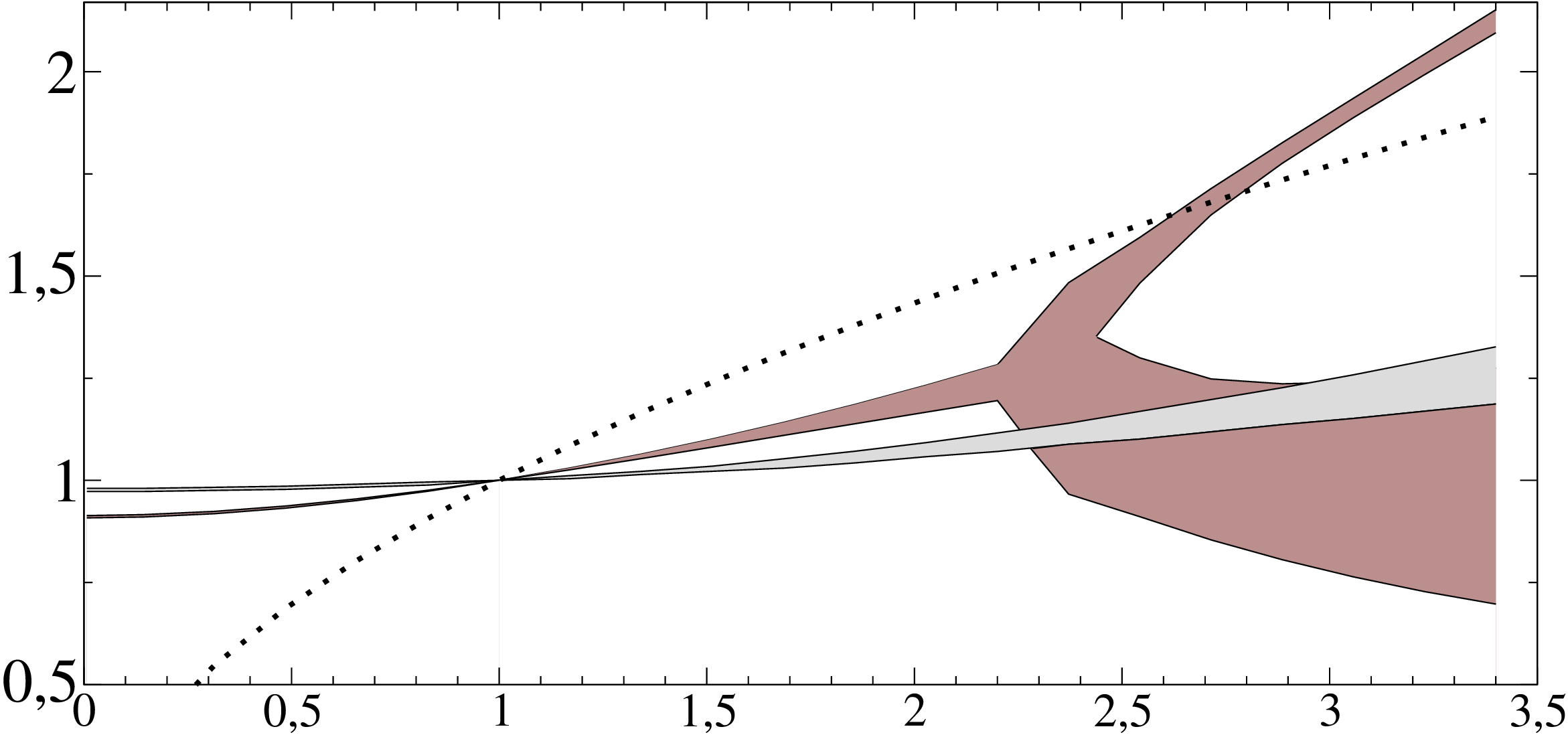}
\includegraphics[scale=0.21,angle=0]{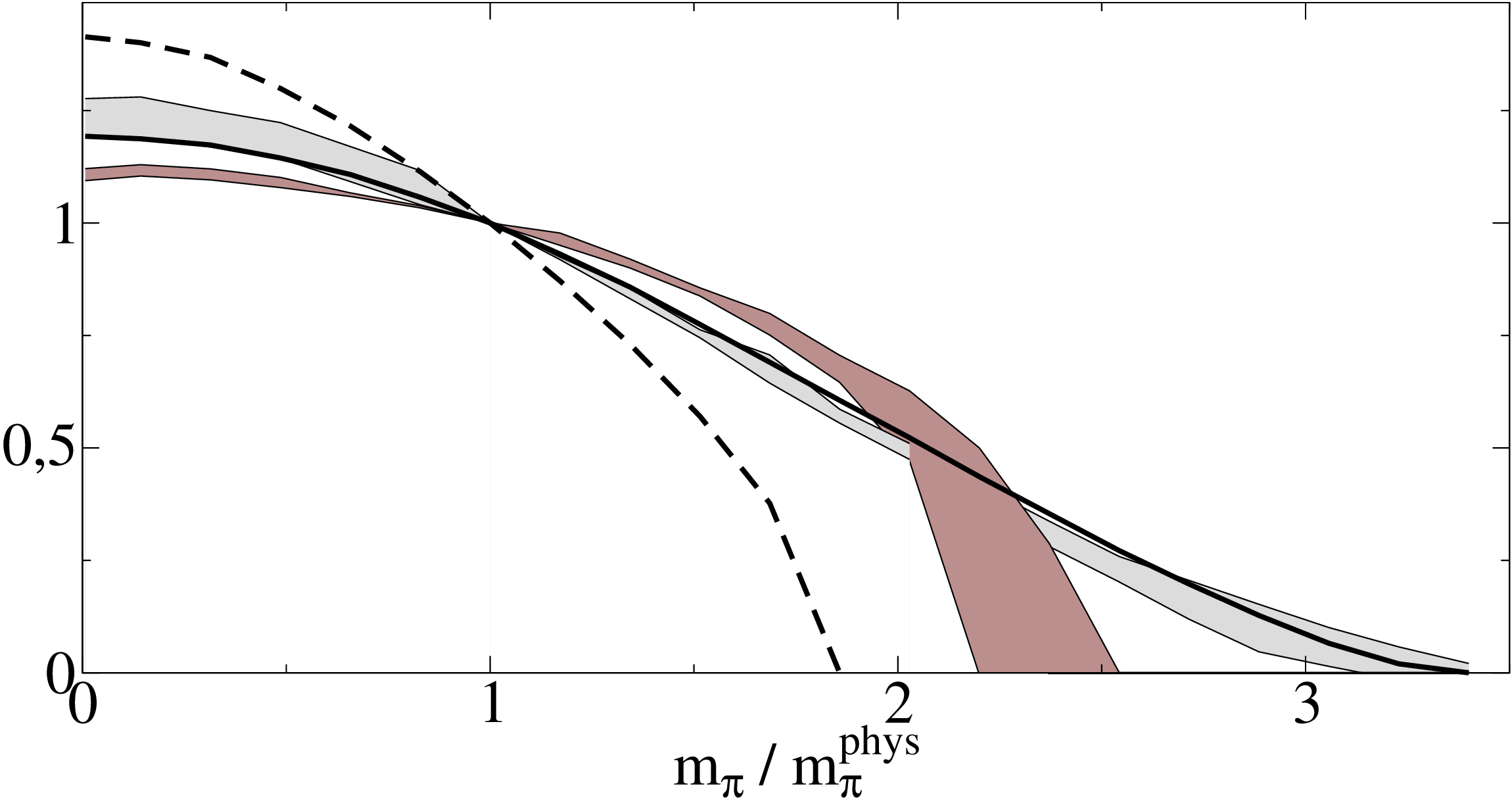}
\caption{ $m_\pi$ dependence of resonance masses (upper panel)
  and widths (lower panel) in units of the physical values. 
 In both panels the dark (light) band shows the results for
the $\sigma$ ($\rho$). The width of the bands reflects the uncertainties
induced from the uncertainties in the LEC. The dotted line shows the $\sigma$ mass dependence
  estimated in Ref.~\cite{Jeltema:1999na}. The dashed (continuous)
line shows the $m_\pi$ dependence of the $\sigma$ ($\rho$) 
width from the change of phase space only, assuming
a constant coupling of the resonance to $\pi\pi$.}
\label{mrymsSplitLsM}
\end{figure}
In the upper panel of Fig.~\ref{mrymsSplitLsM} we show the $m_\pi$
dependence of $M_\sigma$ and $M_\rho$ normalized to their physical
values. The bands cover the LEC uncertainties. Note, that significant,
additional uncertainties may emerge at the two loop level for pion
masses larger than 0.3 GeV --- see, e.g., Ref.~\cite{stephan}.  We
see that both masses grow with increasing $m_\pi$, but the rise of
$M_\sigma$ is stronger than that of $M_\rho$, and again we see that
around $m_\pi\simeq 0.33\,{\rm GeV}$ the $\sigma$ state splits into two
virtual states with different behavior. The upper branch moves closer
to threshold and thus has the biggest influence in the physical
region, eventually jumping to the first Riemann sheet.  Note that the
$m_\pi$ dependence of $M_\sigma$ is much softer than that suggested in
the model of \cite{Jeltema:1999na}, shown as the dotted line, which in
addition does not show the virtual pole splitting.

In the lower panel of Fig.~\ref{mrymsSplitLsM} we show the $m_\pi$
dependence of $\Gamma_\sigma$ and $\Gamma_\rho$ normalized to their
physical values.  The decrease in $\Gamma_\rho$ is largely kinematical,
following remarkably well the expected reduction from phase space as
$m_\pi$ and $M_\rho$ increase. In other words, the effective coupling of the
$\rho$ to $\pi\pi$ is almost $m_\pi$ independent. This was assumed
in the analysis of Ref.~\cite{rhowidthlattice}; however, so far this
assumption has not been supported by theory.  In sharp contrast to this
behavior is the one of $\Gamma_\sigma$.  This suggests a strong
pion mass dependence of the $\sigma$ coupling to two pions,
necessarily present for molecular states~\cite{baru,mol}.

\begin{figure}
\begin{center}
\includegraphics[scale=0.25,angle=0]{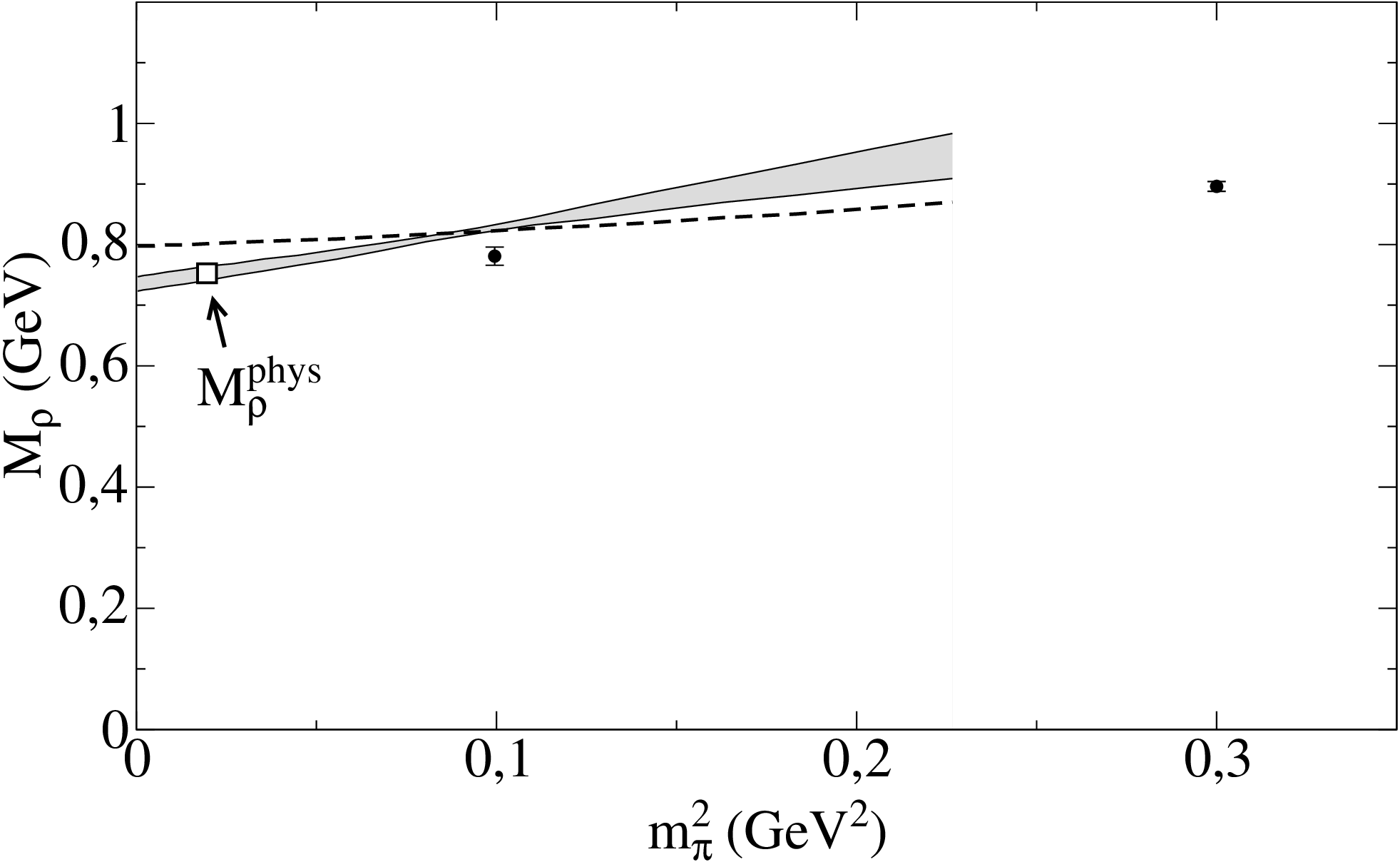}
\caption{The grey band shows the $m_\pi$ dependence of $\rho$ pole
  mass from the IAM versus recent lattice results from
  \cite{Aoki:1999ff}. The dashed line is the IAM result for
$N_c=10$.}
\label{lattice-BN}
\end{center}
\end{figure}

Fig.~\ref{lattice-BN} is a comparison of our results for the $m_\pi$
dependence of $M_\rho$ with some recent unquenched lattice results
\cite{Aoki:1999ff}, which deserves several words of caution.  In
particular, our approach only ensures the $m_\pi$ dependence contained
in the NLO ChPT series --- it is, e.g., lacking terms of order
$m_\pi^2 p^4$ and higher. In addition, $M_\rho$ is the `pole mass'
which, particularly for physical values, is deep in the complex plane,
while, due to the finite lattice volume, the minimum energy with
which pions are produced on the lattice is larger than the resulting
$M_\rho$.  In our formalism, we can mimic a narrow $\rho$ by increasing
the number of colors \cite{Pelaez:2006nj}. We also show the result of
rescaling the IAM ChPT amplitudes from $N_c=3$ to $N_c=10$, which
effectively reduces $\Gamma_\rho$ by a factor of $3/10$.  Although this
narrowing effect is not exactly the same as that on the lattice -- it
is more like quenching the lattice results, since the large $N_c$
expansion actually suppresses quark loops -- it is encouraging that
making the $\rho$ artificially narrower yields a better agreement with
the quark-mass dependence of the lattice data.  With these caveats in
mind our results are in qualitative agreement with the lattice
results.
Following Ref.~\cite{Bruns:2004tj} one may write
$
M_\rho= M_\rho^0+c_1 m_\pi^2+ O(m_\pi^3) \, ,
$
where the $c_i$ parameters are expected to be of order one and $0.65\,
{\rm GeV}\leq M_\rho^0\leq 0.80\, {\rm GeV}$.  This is confirmed by a
fit to lattice data~\cite{Aoki:1999ff}. From our approach we predict
$M_\rho^0=0.735\pm 0.0017\, {\rm GeV}$.  Furthermore, the IAM
reproduces $M_\rho$ at the physical value of $m_\pi$, where higher
orders in the $\rho$ mass formula are $\simeq
15\%$~\cite{Bruns:2004tj}. Within that uncertainty, we thus get the
prediction $c_1=0.90\pm 0.11\pm0.13\,{\rm GeV}^{-1}$.  Although the
quark-mass dependence of our calculation is steeper than that of
Ref.~\cite{Aoki:1999ff}, the extracted values are still consistent
with the expectations mentioned above --- again we remind the reader
that the $m_\pi$ dependence included is correct only to NLO in ChPT.

To summarize, we presented a prediction for the quark mass
dependence of the lightest resonances in QCD, namely the $\rho$ and
the $\sigma$ meson based on chiral perturbation theory at
next--to--leading order together with the inverse amplitude method.
We showed that the mass of the $\rho$ has a very smooth $m_\pi$
dependence and its coupling to $\pi\pi$ is almost quark-mass
independent --- in Ref.~\cite{rhowidthlattice}, this was only assumed
without further evidence. The mass of the $\sigma$, on the other hand, shows
a pronounced non--analyticity when $m_\pi$ is varied. In addition, its
effective coupling to $\pi\pi$ is strongly $m_\pi$ dependent. This
is interpreted as additional evidence for a significant molecular admixture
in the sigma, consistent with previous analyses, and will
be important for future chiral extrapolations of lattice data
for $s$--wave resonances.

We thank G. Colangelo, S. D\"urr, V.V. Flambaum, U.--G. Mei{\ss}ner and A. Rusetsky
for their useful comments.

\end{document}